\documentclass[%
preprint,
 amsmath,amssymb,
 aps,
]{revtex4-1}
\usepackage[normalem]{ulem}
\usepackage{amsmath}
\usepackage{graphicx}% Include figure files
\usepackage{dcolumn}% Align table columns on decimal point
\usepackage{bm}% bold math
\begin{document}

\preprint{APS/123-QED}

\title{The transparency of graphene and other direct-gap two dimensional materials}% Force line breaks with \\
%\thanks{A footnote to the article title}

\author{Daniel J. Merthe}
  %\email{Corresponding author: merthe@usc.edu}
  \email{merthe@usc.edu}
\author{Vitaly V. Kresin}
\email{kresin@usc.edu}

\affiliation{Department of Physics and Astronomy, University of Southern California, Los Angeles, California 90089 , USA}

\date{\today}% It is always \today, today,

\begin{abstract}
Graphene and other two-dimensional materials display remarkable optical properties, including a simple transparency of $T \approx 1 - \pi \alpha$ for visible light. Most theoretical rationalizations of this "universal" opacity employ a model coupling light to the electron's crystal momentum and put emphasis on the linear dispersion of the graphene bands.  However, such a formulation of interband absorption is not allowable within band structure theory, because it conflates the crystal momentum label with the canonical momentum operator.  We show that the physical origin of the optical behavior of graphene can be explained within a straightforward picture with the correct use of canonical momentum coupling.  Its essence lies in the two-dimensional character of the density of states rather than in the precise dispersion relation, and therefore the discussion is applicable to other systems such as semiconductor membranes.  At higher energies the calculation predicts a peak corresponding to a van Hove singularity as well as  a specific asymmetry in the absorption spectrum of graphene, in agreement with previous results.
\end{abstract}

\pacs{Valid PACS appear here}% PACS, the Physics and Astronomy
\maketitle

\section{\label{sec:level1}Introduction}
Continuously improving fabrication and characterization techniques have revealed interesting and elegant properties of graphene and other two-dimensional (2D) materials. The optical properties of graphene \cite{Nair:2008aa, Mak:2008aa, Mak:2011aa, Mak:2012aa} as well as membranes of InAs \cite{Fang:2013aa} have been measured in recent studies. Remarkably, the transparency of both materials in the visible range is simply $T \approx 1-\pi \alpha$, where $\alpha = e^2 / \hbar c$ is the fine structure constant. This suggests that the optical conductivity of 2D semiconductors and semimetals possesses certain universal features.

The opacity of graphene due to interband transitions has been calculated analytically by a number of authors \cite{Pedersen:2003aa,Katsnelson:2007aa, Falkovsky:2008aa, Pedersen:2009aa, Mecklenburg:2010aa, Kravets:2010aa,Bacsi:2013aa}. The majority of these treatments proceed by writing down the tight-binding Hamiltonian in the pseudo-relativistic form, $\hat{H} = v_F \hat{\mathbf{\sigma}} \cdot \hbar \mathbf{k}$, where $\hat{\mathbf{\sigma}}$ is the Pauli vector and $\mathbf{k}$ is the in-plane crystal momentum of the basis states, and expressing the electron's coupling to the electromagnetic field via the standard vector potential substitution (sometimes called ``minimal coupling") $\hbar \mathbf{k} \rightarrow \hbar \mathbf{k}+e \mathbf{A}/c$. In this presentation, the linear Dirac-cone dispersion of graphene is viewed as a key element underlying its optical properties.

However, the above procedure has a fundamental conceptual flaw\footnote{An elementary argument illustrates that the $\mathbf{k}$-substitution procedure is fundamentally inconsistent. In the mentioned pseudo-relativistic tight-binding Hamiltonian, $\hat{H} = v_F \hat{\mathbf{\sigma}} \cdot \hbar \mathbf{k}$, only $\hat{\mathbf{\sigma}}$ is an operator while  $\mathbf{k}$ is simply a vector.  Therefore in the eigenfunction (i.e., band structure) basis, the Hamiltonian becomes diagonal ($\hat{H} = v_F \sigma_z \hbar k$).  Now a  $\mathbf{k}$-substitution yields a completely diagonal perturbation Hamiltonian, meaning that no absorption can take place. However, in a proper quantum formulation, a similarity transformation cannot affect the physical description.}. Specifically, $\hbar \mathbf{k}$ is not the momentum of the electron but the crystal momentum, i.e. a state label (quantum number) rather than an operator. The actual full non-relativistic graphene Hamiltonian is quadratic in the momentum operator $\hat{\mathbf{p}}$, and it is the latter which must be augmented with the vector potential.

It is indeed possible to describe the evolution of the Bloch states of carriers within a single band $n$ under the influence of an external DC field $\hat{H}'$ by an effective Hamiltonian $\hat{H}_{eff}=E_n(\hat{\mathbf{p}}/\hbar)+\hat{H}'$ where $E_n(\mathbf{k})$ is the band energy \cite{Economou:2010aa}. This formalism underlies the well-known semiclassical equations of carrier transport.  However, it is strictly limited to intraband dynamics and there is absolutely no justification for applying such a procedure to interband transitions induced by AC fields. Indeed, one finds that every textbook discussion of interband absorption necessarily reverts to matrix elements of $\hat{\mathbf{p}}$ \cite{Ashcroft:1976aa, Kittel:1987aa, Snoke:2009aa}.

The aforementioned finding \cite{Fang:2013aa} that a 2D nanomembrane of a direct-gap semiconductor also displays the same $\pi \alpha$ quantum of absorption attests that it is the dimensionality of these systems (specifically, the electronic density of states in 2D) that underlies their beautiful optical characteristics. In this paper we show that by applying the "minimal coupling" substitution to the canonical momentum $\hat{\mathbf{p}}$ and using Fermi's Golden Rule, one can straightforwardly reproduce the experimentally observed constant opacity for low frequencies and also the qualitative features of graphene's absorption spectrum for higher frequencies, which are noticeably different from what one obtains by using a "$\mathbf{k}$-substitution".

\section{Low Frequency Light Absorption by 2D Electrons}
Only general considerations are needed to characterize the low frequency absorption rate of electrons in 2D systems with a conical or higher power-law band structure. Introducing an electromagnetic field invokes the substitution $\hat{\mathbf{p}} \rightarrow \hat{\mathbf{p}}+e\mathbf{A}/c$, rendering a perturbation Hamiltonian, $\hat{H'} = (e/mc)\mathbf{A} \cdot \hat{\mathbf{p}}$, where $-e$ and $m$ are the charge and mass of the electron. We shall consider a normally incident circularly polarized plane wave, which in the $z=0$ plane takes the form,
\begin{equation}\label{eq:emwave}
\mathbf{A}(z=0,t) = (c \mathcal{E} / 2 \omega_0) (\hat{\mathbf{x}}+i\hat{\mathbf{y}}) e^{-i\omega_0 t}
\end{equation}
where $c$ is the speed of light and $\mathcal{E}$ is the electric field amplitude. At sufficiently low temperature, only vertical transitions $\Delta \mathbf{k} = 0$ are allowed, and these interband transitions dominate the absorption process. Then, according to Fermi's Golden Rule, the rate of vertical transitions from the valence band ("-") to the conduction band ("+") at a point $\mathbf{k}$ in the Brillouin Zone (BZ) under such a perturbation is given by
\begin{equation}\label{eq:FermiGR}
R(\mathbf{k}) = \frac{\pi e^2 \mathcal{E}^2}{2 \hbar m^2 \omega_0^2} ||\langle +,\mathbf{k} |\hat{\mathbf{p}}| -, \mathbf{k} \rangle ||^2 \delta(E_+ (\mathbf{k}) - E_- (\mathbf{k}) - \hbar \omega_0)
\end{equation}

Consider a 2D direct gap material with a dispersion relation of the form
\begin{equation}\label{eq:powerE_k}
E_{\pm}(\mathbf{k}) = \pm \left[ \frac{\Delta}{2} + C_{\pm} \ q^\lambda \right]
\end{equation}
where $q\equiv |\mathbf{k-K}|$ is the distance in $k$-space from the band gap minimum, $E_{+}(\mathbf{K})-E_{-}(\mathbf{K})=\Delta$, $C_{\pm}$ and $\lambda$ are material dependent parameters, and the positive (negative) sign corresponds to the conduction (valence) bands. Graphene, for example, corresponds to $C_{\pm} = \pm \hbar v_f $, $\Delta \rightarrow 0$ and $\lambda \rightarrow 1$.

To calculate the momentum matrix element for low energy excitations we use a well known result of $\mathbf{k\cdot p}$ perturbation theory. This calculation is similar to that of Fang et. al \cite{Fang:2013aa}.  The effective mass in the $n^{th}$ band is given by \cite{Ashcroft:1976aa}
\begin{equation}
\frac{1}{m_{n} (\mathbf{k})} = \frac{1}{m} + \frac{2}{m^2} \sum_{n'\neq n} \frac{||\langle n, \mathbf{k} | \hat{\mathbf{p}}| n', \mathbf{k} \rangle ||^2}{E_{n}(\mathbf{k})-E_{n'}(\mathbf{k})}
\end{equation}
In the limit of small $q$ and $\Delta$, the small energy difference between the valence and conduction bands causes the term coupling these two bands to dominate the sum. In this approximation, we obtain
\begin{equation}\label{eq:OneTermMass}
\frac{1}{m_{+} (\mathbf{k})} +  \frac{1}{m_{-} (\mathbf{k})} \approx \frac{4}{m^2} \frac{||\langle +, \mathbf{k} | \hat{\mathbf{p}}| -, \mathbf{k} \rangle ||^2}{E_{+}(\mathbf{k})-E_{-}(\mathbf{k})}
\end{equation}
where $m_+$ ($m_-$) is the effective mass of electrons (holes) in the conduction (valence) band. Using the dispersion relation in Eq. \eqref{eq:powerE_k} to obtain the effective masses and rearranging Eq. \eqref{eq:OneTermMass}, the modulus square of the momentum matrix element is
\begin{equation}\label{eq:pmatrix1}
||\langle +, \mathbf{k} | \hat{\mathbf{p}}| -, \mathbf{k} \rangle ||^2 = \frac{\lambda^2 m^2}{4 \hbar^2} \left[  \Delta + (C_+ + C_-)q^\lambda\right] ( C_+ + C_- ) q^{\lambda-2}
\end{equation}

The total transition rate is given by integrating Eq. \eqref{eq:FermiGR} over the Brillouin zone. In order to transform this integration over $\mathbf{k}$ into an integration over energy difference,
\begin{equation}\label{eq:energydifference}
\epsilon(\mathbf{k}) \equiv E_{+}(\mathbf{k})-E_{-}(\mathbf{k}) = \Delta + (C_+ + C_-)q^\lambda
\end{equation}
the joint density of states $\rho(\epsilon) $ is defined by the relation,
\begin{equation}
\rho(\epsilon)\ = \frac{2gS}{4\pi^2} \int d^2 k \ \delta(E_+ (\mathbf{k}) - E_- (\mathbf{k}) - \epsilon)
\end{equation}
where $S$ is the area of the material surface and $g$ is the valley degeneracy factor (the total angle around the band extrema divided by $2\pi$). Using the transformation of delta functions upon change of variables, from this and Eq. \eqref{eq:powerE_k} we obtain,
\begin{equation}
\rho(\epsilon) = \frac{gS}{\pi \lambda (C_+ + C_-)} \left( \frac{C_+ + C_-}{\epsilon - \Delta}\right)^{1 - \frac{2}{\lambda}} \Theta (\epsilon - \Delta)
\end{equation}
where $\Theta$ is the step function. Combining this with Eqs. \eqref{eq:FermiGR} and \eqref{eq:pmatrix1}, the total rate of transitions is
\begin{equation}\label{eq:genrate}
\begin{array}{rl}
R & = \int d\epsilon  \ \frac{\pi e^2 \mathcal{E}^2}{2 \hbar m^2 \omega_0^2} ||\langle +,\mathbf{k} |\hat{\mathbf{p}}| -, \mathbf{k} \rangle ||^2 \rho(\epsilon) \delta(\epsilon- \hbar \omega_0) \\
    & = \frac{g\lambda \pi}{2} \frac{e^2}{\hbar c} \frac{cS \mathcal{E}^2}{4\pi \hbar \omega_0} \Theta(\hbar \omega_0 - \Delta) \\
    & = \frac{g\lambda}{2} \pi \alpha R_0 \ \Theta(\hbar \omega_0 - \Delta)
\end{array}
\end{equation}
where $\alpha = e^2/\hbar c$ is the fine structure constant and $R_0 = c S \mathcal{E}^2/4 \pi \hbar\omega_0$ is the incidence rate of photons. In going from the first to the second equality, Eq. \eqref{eq:energydifference} is used to express the modulus square of the momentum matrix element in terms of $\epsilon$. Remarkably, the material dependent parameters $C_\pm$ cancel out between the density of states and the momentum matrix element. However, the power parameter $\lambda$ still affects the total transition rate.

For a semiconductor membrane with a band gap at the center of the BZ, $\lambda=2$ and $g=1$, we find an opacity of
\begin{equation}
R/R_0 = \pi \alpha
\end{equation}
for frequencies above the band gap. For graphene with its band gaps at the corners, we have $\lambda = 1$ and $g = 2$ (six corners, each contributing an angle of $2\pi/3$; see Figs. 1b and 2), and we obtain from  Eq. \eqref{eq:genrate} exactly the same value. This value, instead of being "universal" as suggested in Ref. \cite{Fang:2013aa}, should more appropriately be viewed as a numerical confluence.

For multilayers of ABC-stacked graphene, the electronic energy eigenvalues near the Fermi level can be written in the form,
\begin{equation}\label{eq:NlayerlowEHam}
E_{\pm}(\mathbf{k}) \propto \pm q^N
\end{equation}
where $N$ is the number of layers \cite{Min:2008aa}. Thus, with $g = 2$ and $\lambda = N$, Eq. \eqref{eq:genrate} yields a low frequency opacity of
\begin{equation}
R/R_0 = N \pi \alpha
\end{equation}
which is in good agreement with the observations in Ref. \cite{Mak:2011aa} for $N$ up to 4.

Thus based on the 2D character of the electron system and using only a general dispersion relation, we are able to explain in a correct quantum band structure framework the opacity of monolayer graphene, multilayer graphene, and direct gap semiconductor membranes \cite{Nair:2008aa, Mak:2008aa, Mak:2012aa, Fang:2013aa}. However, this calculation is only valid in the limit $E_{+}(\mathbf{k})-E_{-}(\mathbf{k}) \rightarrow 0$. In the case of graphene, the absorption rate can be readily evaluated for higher electromagnetic frequencies within the tight-binding approximation. This more complete calculation also avoids questions which may arise regarding the $\Delta, q \rightarrow 0$ limit in view of Eq. \eqref{eq:OneTermMass}. 

\section{Frequency Dependent Absorption in the Graphene Structure}
In order to calculate the transition rate for a broader frequency range, the momentum matrix elements $\langle \pm,\mathbf{k} |\hat{\mathbf{p}}| -, \mathbf{k} \rangle$ need to be evaluated explicitly.

We proceed in the H\"{u}ckel tight-binding approximation as usual, using linear combinations of $2p_z$ orbitals at each carbon atom in the lattice as a basis set for the Bloch wave functions. For reference, the lattice geometry is shown in Fig 1a. Through the standard analysis \cite{Castro-Neto:2009aa}, one finds that the energy eigenvalues of the valence and conduction bands are given in the nearest-neighbor approximation by

\begin{equation}\label{eq:energy}
\begin{array}{c}
E_{\pm}(\mathbf{k}) = \pm \gamma_0 | f(\mathbf{k})| \\
f(\mathbf{k}) = 1 + e^{-i\mathbf{k}\cdot \mathbf{a}_1}+ e^{-i\mathbf{k}\cdot \mathbf{a}_2}
\end{array}
\end{equation}
with "hopping" energy $\gamma_0 \approx 2.8$ eV and the upper (lower) sign indicating the conductance (valence) band, as above. These energy bands are plotted in Fig. 1b. The corresponding eigenfunctions are

\begin{equation}
\begin{array}{rl}
\psi_{\pm, \mathbf{k}} (\mathbf{x}) = \frac{1}{\sqrt{2 N}} \sum_{\mathbf{R}} & e^{i \mathbf{k}\cdot\mathbf{R}} [ \chi(\mathbf{x-R}) \\
& \pm e^{i \phi} \chi(\mathbf{x-R-h})]
\end{array}
\end{equation}
where $\chi(\mathbf{x})$ is the carbon $2p_z$ orbital, the phase factor is $e^{i \phi(\mathbf{k})} \equiv f(\mathbf{k})/ |f(\mathbf{k})|$, and the sum is over all Bravais lattice points $\mathbf{R}$ (sublattice A in Fig. 1a).

\begin{figure}
\centering
\includegraphics[width=3.34in]{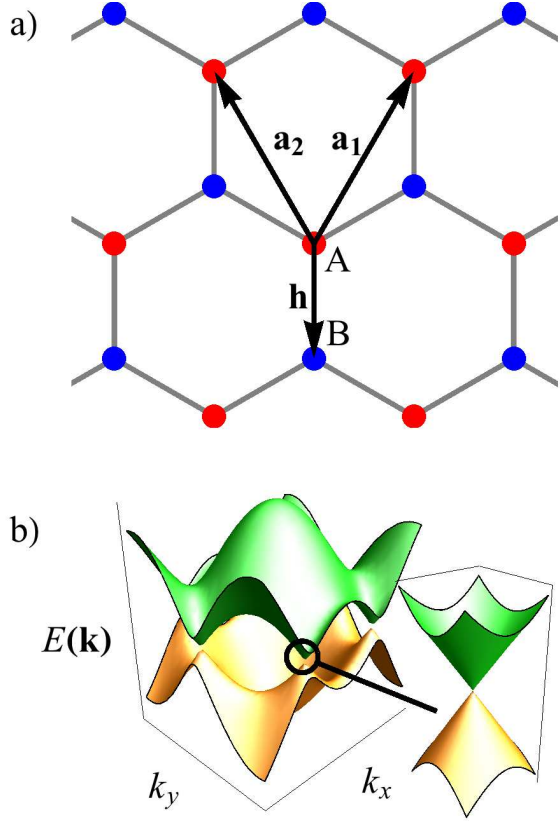}
\caption{(color online). (a) Lattice geometry of graphene and (b) band structure in the tight binding approximation, with conical dispersion near the Dirac points (circled).}
\end{figure}

For the diagonal (intraband) momentum matrix elements we already have the relation,

\begin{equation}
\frac{1}{m} \langle -,\mathbf{k} |\hat{\mathbf{p}}| -, \mathbf{k} \rangle = \frac{1}{\hbar} \frac{\partial E_-(\mathbf{k})}{\partial \mathbf{k}}  \equiv \bar{\mathbf{v}}_-(\mathbf{k})
\end{equation}
for the group velocity $\bar{\mathbf{v}}_-(\mathbf{k}) $ of electrons or holes in the valence band \cite{Ashcroft:1976aa}, but as will be seen below it is useful to calculate this matrix element simultaneously. Generally, the matrix elements are given by

\begin{equation}\label{momentumelem}
\begin{array}{rl}
\langle \pm,\mathbf{k} |\hat{\mathbf{p}}| -, \mathbf{k} \rangle = & \frac{1}{2N} \sum_{\mathbf{R,R}'} e^{i\mathbf{k}\cdot(\mathbf{R-R}')} \\
& \times [ \int d^3 x \ \chi(\mathbf{x-R}') \hat{\mathbf{p}} \chi(\mathbf{x-R}) \\
& - e^{i \phi} \int d^3 x \ \chi(\mathbf{x-R}') \hat{\mathbf{p}} \chi(\mathbf{x-R-h}) \\
& \pm e^{-i \phi} \int d^3 x \ \chi(\mathbf{x-R'-h}) \hat{\mathbf{p}} \chi(\mathbf{x-R}) \\
&  \mp \int d^3 x \ \chi(\mathbf{x-R'-h}) \hat{\mathbf{p}} \chi(\mathbf{x-R-h}) ]
\end{array}
\end{equation}

 The first and last integrals vanish:  for $\mathbf{R} \neq \mathbf{R}'$ due to the assumption of wavefunction overlap only between nearest neighbors, and  for $\mathbf{R}=\mathbf{R}'$ due to the parity symmetry of $\chi(\mathbf{x})$. Likewise assuming only nearest neighbor overlap, the remaining two integrals reduce to
\begin{equation}\label{eq:momoverlap1}
\begin{array}{rl}
\int d^3 x & \chi(\mathbf{x-R}') \hat{\mathbf{p}} \chi(\mathbf{x-R-h}) \\
		& = \delta_{\mathbf{R,R'}} \int d^3 x \ \chi(\mathbf{x}) \hat{\mathbf{p}} \chi(\mathbf{x-h}) \\
		& + \delta_{\mathbf{R,R'+a}_1} \int d^3 x \ \chi(\mathbf{x}) \hat{\mathbf{p}} \chi(\mathbf{x-h-a}_1) \\
		& + \delta_{\mathbf{R,R'+a}_2} \int d^3 x \ \chi(\mathbf{x}) \hat{\mathbf{p}} \chi(\mathbf{x-h-a}_2)
\end{array}
\end{equation}
\begin{equation}\label{eq:momoverlap2}
\begin{array}{rl}
\int d^3 x & \chi(\mathbf{x-R'-h}) \hat{\mathbf{p}} \chi(\mathbf{x+R}) \\
		& = \delta_{\mathbf{R,R'}} \int d^3 x \ \chi(\mathbf{x}) \hat{\mathbf{p}} \chi(\mathbf{x+h}) \\
		& + \delta_{\mathbf{R,R'-a}_1} \int d^3 x \ \chi(\mathbf{x}) \hat{\mathbf{p}} \chi(\mathbf{x+h+a}_1) \\
		& + \delta_{\mathbf{R,R'-a}_2} \int d^3 x \ \chi(\mathbf{x}) \hat{\mathbf{p}} \chi(\mathbf{x+h+a}_2)
\end{array}
\end{equation}
where $\delta_{\mathbf{w},\mathbf{w'}}$ is the Kronecker symbol. These integrals can be further simplified with the following consideration. Let $\mathbf{s}=s\hat{\mathbf{u}}$ be a displacement in the $xy$ plane, where $\hat{\mathbf{u}}$ is a unit vector. Then,
\begin{equation}\label{eq:overlapp}
\begin{array}{rl}
\int d^3 x \ \chi(\mathbf{x}) \hat{\mathbf{p}} \chi(\mathbf{x-s}) &= -i\hbar \int d^3 x \ \chi(\mathbf{x}) \frac{\partial}{\partial \mathbf{x}} \chi(\mathbf{x-s}) \\
										&= i\hbar  \frac{\partial}{\partial \mathbf{s}} \int d^3 x \  \chi(\mathbf{x}) \chi(\mathbf{x-s}) \\
\end{array}
\end{equation}
By the azimuthal symmetry of $\chi(\mathbf{x})$, this latter overlap integral is independent of the direction of $\mathbf{s}$, and is a function $F(s)$ of only its magnitude $s$,
\begin{equation}\label{eq:overlapintegral}
\int d^3 x \ \chi(\mathbf{x}) \chi(\mathbf{x-s})=\int d^3 x \ \chi(\mathbf{x}) \chi(\mathbf{x}-s\hat{\mathbf{u}}) \equiv F(s)
\end{equation}
Therefore, defining $D(s) \equiv (\partial / \partial s) F(s)$, the integral in Eq. \eqref{eq:overlapp} is
\begin{equation}\label{eq:momoverlap}
\int d^3 x \chi(\mathbf{x}) \hat{\mathbf{p}} \chi(\mathbf{x-s}) = i\hbar \hat{\mathbf{u}} D(s)
\end{equation}
With this applied to Eqs. \eqref{eq:momoverlap1} and \eqref{eq:momoverlap2} and the nearest-neighbor separation $a$, Eq. \eqref{momentumelem} can be rewritten as
\begin{equation}\label{eq:diagelem}
\begin{array}{rcl}
\langle -,\mathbf{k} |\hat{\mathbf{p}}| -, \mathbf{k} \rangle =& \hbar D(a)  [&\frac{\mathbf{h}}{a} \sin(\phi(\mathbf{k})) \\
													&&+ \frac{\mathbf{h+a}_1}{a} \sin(\phi(\mathbf{k}) + \mathbf{k}\cdot\mathbf{a}_1) \\
													&& + \frac{\mathbf{h+a}_2}{a} \sin(\phi(\mathbf{k}) + \mathbf{k}\cdot\mathbf{a}_2)] \\
\end{array}
\end{equation}
and
\begin{equation}\label{eq:offdiagelem}
\begin{array}{rcl}
\langle +,\mathbf{k} |\hat{\mathbf{p}}| -, \mathbf{k} \rangle =& -i \hbar D(a)  [&\frac{\mathbf{h}}{a} \cos(\phi(\mathbf{k})) \\
													&&+ \frac{\mathbf{h+a}_1}{a} \cos(\phi(\mathbf{k}) + \mathbf{k}\cdot\mathbf{a}_1) \\
													&& + \frac{\mathbf{h+a}_2}{a} \cos(\phi(\mathbf{k}) + \mathbf{k}\cdot\mathbf{a}_2)] \\
\end{array}
\end{equation}
The quantity $D(a)$ can in principle be evaluated by explicitly computing the integral in Eq. \eqref{eq:overlapintegral} and taking the derivative. However, with knowledge of $\langle -,\mathbf{k} |\hat{\mathbf{p}}| -, \mathbf{k} \rangle = m \bar{\mathbf{v}}_-(\mathbf{k}) $ from the band structure, $D(a)$ can be eliminated using Eq. \eqref{eq:diagelem}. Specifically, near each of the Dirac points $\mathbf{K}_D$ where the dispersion becomes conical, we have
\begin{equation}
\bar{\mathbf{v}}_-(\mathbf{k}\rightarrow\mathbf{K}_D) = -v_F \frac{\mathbf{k-K}_D}{|\mathbf{k-K}_D|} = \frac{3\hbar D(a)}{2m} \frac{\mathbf{k-K}_D}{|\mathbf{k-K}_D|}
 \end{equation}
where $v_F$ is the Fermi velocity and the expression on the right comes from expanding Eq. \eqref{eq:diagelem} about the point $\mathbf{k} = \mathbf{K}_D$. Therefore, $D(a) = -2mv_F/3\hbar$ and the off-diagonal momentum matrix element is
\begin{widetext}
\begin{equation}\label{eq:mtrxelem}
\langle +,\mathbf{k} |\hat{\mathbf{p}}| -, \mathbf{k} \rangle = \frac{2i m v_F}{3} \left[\frac{\mathbf{h}}{a} \cos(\phi(\mathbf{k}))+ \frac{\mathbf{h+a}_1}{a} \cos(\phi(\mathbf{k}) + \mathbf{k}\cdot\mathbf{a}_1)+ \frac{\mathbf{h+a}_2}{a} \cos(\phi(\mathbf{k}) + \mathbf{k}\cdot\mathbf{a}_2)\right]
\end{equation}
\end{widetext}
The norm of this matrix element as a function of $\mathbf{k}$ is plotted in Fig. 2. The Dirac points are saddle points, where the norm has the value $m v_F$, with the global maxima occurring at the midpoints between Dirac points, where the norm is $2 m v_F$.
\begin{figure}
\centering
\includegraphics[width=3.34in]{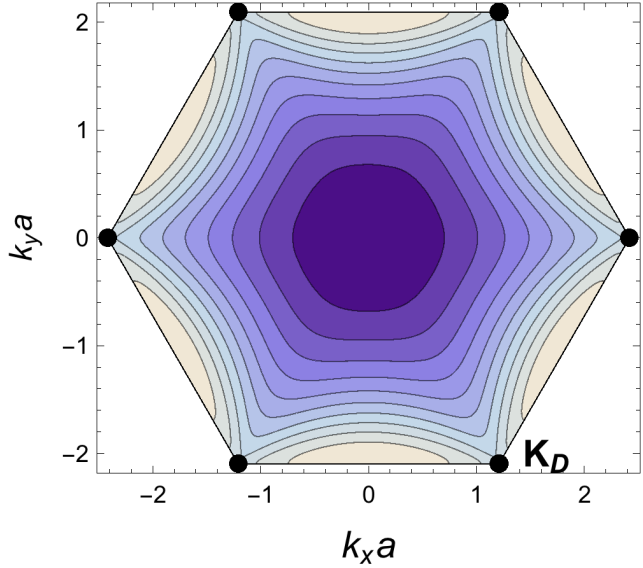}
\caption{(color online). Norm of the momentum matrix element $ \langle +,\mathbf{k} |\hat{\mathbf{p}}| -, \mathbf{k} \rangle $ as function of $\mathbf{k}$ in the BZ. The black dots are the Dirac points $\mathbf{K}_D$. Lighter (darker) shades indicates higher (lower) values, going to zero at $\mathbf{k} = \mathbf{0}$.}
\end{figure}
Using the expression for the momentum matrix element in Eq. \eqref{eq:mtrxelem}, the opacity can be found by integrating Eq. (\ref{eq:FermiGR}) over the entire BZ for a given light frequency $\omega_0$. This is straightforward near the Dirac points where the constant-energy contours are circular, but becomes less manageable farther away. Therefore, we performed the integration by the Monte Carlo method, using a narrow rectangular gate in place of the delta function. The solid red curve in the lower panel of Fig. 3 shows the calculated opacity.

The dotted blue curve in the lower panel of Fig. 3 shows the opacity calculated via a $\mathbf{k}$-substitution applied to the tight-binding Hamiltonian from Ref. \cite{Semenoff:1984aa}. The two curves are qualitatively similar, with a peak near 5.6 eV due to the high joint density of states at this energy (as shown in the upper panel of Fig. 3). At low frequencies both curves show that the opacity rises gradually from $\pi \alpha$, closely matching the results of Refs. \cite{Nair:2008aa, Mak:2011aa, Mak:2012aa}. However, as the frequency increases the two curves diverge noticeably. With the $\mathbf{k}$-substitution shortcut the opacity levels off near $\pi \alpha$ after the peak in a symmetric manner.  But the present calculation based on the full evaluation of the matrix elements of  $\mathbf{p}$ shows the opacity decreasing for high frequencies, and is more asymmetric as a result.  In this, it has a closer similarity to the experimental profile in Refs. \cite{Mak:2011aa,Mak:2012aa}, which was also observed to have a high degree of asymmetry about the maximum, see Fig. 3. (The experimental peak was found to be red-shifted to 4.62 eV; this behavior was assigned to the influence of excitonic effects \cite{Kravets:2010aa,Mak:2012aa}. The dip in the data for energies $<$0.5 eV was attributed  \cite{Mak:2012aa} to unintentional doping of the graphene sheets.)

Thus a correct calculation of the optical transparency of graphene, performed by applying the vector potential coupling to the electrons' canonical momentum, is straightforward and produces results which are in good agreement with the available experimental data. The two-dimensional character of the system, rather than the particular dispersion relation, is the dominant factor responsible for the low-frequency "universality" of a constant opacity.  It is straightforward to extend the calculation to doped graphene \cite{Mak:2012aa}, and to apply it to other 2D direct-gap materials.  It would be very interesting if the measurement range of graphene's optical transmission could be extended to higher frequencies, so as to explore its behavior and shape as it dips back below the low-frequency value. This could perhaps be accomplished at a synchrotron facility.

\begin{figure}
\centering
\includegraphics[width=3.34in]{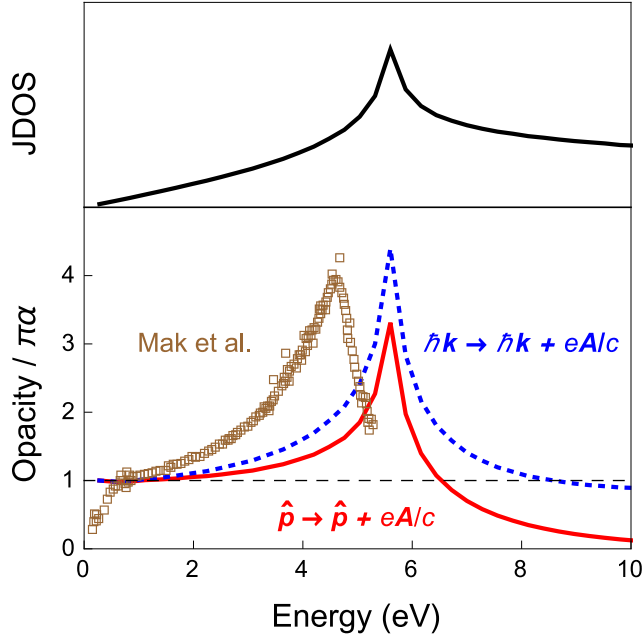}
\caption{(color online). Opacity of graphene calculated with "$\mathbf{p}$-substitution" (solid red curve, lower panel) and "$\mathbf{k}$-substitution" (dotted blue curve) as a function of photon energy, and the calculated joint density of states (JDOS) for valence-conductance band transitions as a function of energy difference. For comparison, the measured opacity of graphene from Ref. \cite{Mak:2012aa} is also shown with brown squares.}
\end{figure}

\section{Acknowledgments}
We would like to thank Prof. Dmitry Budker for stimulating discussions and for bringing this problem to our attention, and Prof. Gerd Bergmann for valuable discussions. This research was supported by the U.S. National Science Foundation under grant CHE-1213410.

\end{document}